\documentclass[structabstract]{aa}  
\usepackage{graphicx}
\usepackage{txfonts}
\def\Mdot {$\dot M$\,}

\def\simlt{\lower.5ex\hbox{$\; \buildrel < \over \sim \;$}}
\begin{document}
\title{UV diagnostic of porosity-free mass-loss estimates in B stars}
\subtitle{}
   \author{R. K. Prinja
           \inst{1} 
           \and
           D. L. Massa \inst{2}
}

   \institute{Department of Physics \& Astronomy, UCL, Gower Street, 
              London
              WC1E 6BT, UK\\
             \email{rkp@star.ucl.ac.uk}
        \and Space Science Institute (SSI), 4750 Walnut St.,
              Boulder, Colorado 80301, USA\\
	     \email{dmassa@SpaceScience.org}
	}

\date{Received April, 2013; accepted 2013}


\abstract
{}
{We seek to establish evidence in UV P~Cygni line profiles that the signs
of wind clumping and porosity vary with velocity. We aim to demonstrate
empirically that while at most wind velocities optically thick clumps cover
only a fraction of the stellar surface, close to the terminal velocity
($v_\infty$) where narrow absorption components (NACs) appear in UV lines
the covering factor is approximately unity.}  {SEI line-synthesis models
are used to determine the radial optical depths ($\tau_{\rm{rad}}(w)$) of
blue and red components of the Si{\sc iv} $\lambda\lambda$1400 resonance
line doublet in a sample of 12 B0 to B4 supergiants. We focus on stars with
well developed NACs and relatively low $v_\infty$ so that the Si{\sc iv}
doublet components can be treated as radiatively decoupled and formed
independently.}  {For all 12 stars the mean optical depth ratio
of the blue to red components is closer to $\sim$ 2 (i.e. the
ratio of oscillator strengths) in the NACs than at intermediate and lower
velocities.  The product of mass-loss rate and ion fraction
(\Mdot\,$q(Si^{3+})$) calculated from the NAC optical depths is a factor
of $\sim$ 2 to 9 higher compared to mass-loss values sampled at $\sim$ 0.4
to 0.6 $v_\infty$.}  {Since the wind effectively becomes `smooth' at the
high NAC velocities and the column density is uniformly distributed over
the stellar disk, the optical depths of the NACs are not seriously affected
by porosity and this feature thus provides the most reliable measurement of
mass-loss rate in the UV lines. Applications of this result to the
weak-wind problem of late O-dwarf stars and the ``P~{\sc v} mass loss
discordance'' in early O supergiants are discussed.}


 \titlerunning{Wind clumping, porosity and mass-loss in B supergiants}
 \authorrunning{R.K. Prinja and D.L. Massa}

\maketitle

\section{Introduction}
The winds of massive stars power and enrich the interstellar medium,
control the evolution of the stars, determine their ultimate fate and the
nature of their remnants, determine the appearance of the
Hertzsprung-Russell diagram of young, massive clusters and star-bursts, and
play a major role in the initial stages of massive star cluster formation
and their subsequent evolution.  The sensitivity of these effects to
stellar winds is especially important in view of recent developments which
suggest that the mass loss rates of OB stars may be in question at the
{\it order-of-magnitude} level, and not just a few $\%$ (e.g. Massa et al.\
2003, Repolust et al. 2004; Bouret et al. 2005; Prinja et al. 2005;
Fullerton et al. 2006; Puls et al. 2006; Cohen et al. 2006).
Investigations of clumping and structures in the stellar winds are pivotal
in understanding how large discrepancies between different mass loss rate
determinations can arise. There are three traditional mass loss diagnostics
for OB stars; all of which assume a homogeneous, spherically symmetric wind
with a monotonic velocity law. These are the radio free-free emission
(typically arising far out in the wind), H$\alpha$ emission (primarily
located near the star) and the UV wind lines (which sample the entire
wind).  In an ideal scenario all three diagnostics would agree.  However,
strong clumping in the winds can cause them to deviate from one another.
Clumping can cause H$\alpha$ and radio observations to {\it overestimate}
mass loss rates, since both of these diagnostics contain density-squared terms.
At the same time, clumping can cause the UV line absorption to {\it
underestimate} mass loss (e.g. Oskinova et al. 2007).

In the present study we re-examine the effects of clumping in the
measurements of mass-loss rates derived from UV resonance line profiles.
Clumping leads to porosity in both velocity and space
\footnote{Throughout this paper we use the terms porous and porosity to
describe the nature of a surface of constant line of sight velocity.  This
surface need not be simply connected spatially.  For example, dense and
rarefied wind structures moving with different velocities may be connected
spatially but lie on different constant velocity surfaces, a property often
referred to as vorocity.}
(e.g. Oskinova et al. 2007; Sundqvist et al. 2010, \v{S}urlan et al.,
2012) i.e. irregularities in the opacity of the wind material that covers
the stellar disk at a certain velocity and accounts for the absorption that
we measure. These irregularities will allow channels of reduced opacity (or
increased transmission) to be present and thus distort our interpretation
of the line-synthesis modelling of the UV absorption troughs.  An observed
wind-formed UV resonance line can be either decoded using a normal
('smooth') line opacity and a small mass-loss rate or a small ('porosity')
line opacity and a higher mass-loss rate. In this paper we argue that the
optical depths of the Narrow Absorption Components (NACs) commonly
seen in OB stars provide the most reliable (i.e.
porosity-free) measure of mass-loss rate in UV resonance lines.

\subsection{Narrow absorption components (NACs)}
Time-series UV studies of Galactic OB stars have repeatedly demonstrated
that wind line profile variability may often be interpreted in terms of
blueward-migrating discrete absorption components (DACs), which recur on
time scales of days, and which narrow (in velocity space) as they
accelerate towards the shortward profile edge (see e.g. Kaper et al. 1996;
Prinja et al. 2002).  Since the timescale of this UV line profile
variability is (in a few case studies) cyclical and commensurate with the
(maximum) rotation period of the star, this suggests an origin that is
somehow connected to the stellar surface. A promising explanation for the
observed cyclical variability is provided by the so-called co-rotating interaction
regions (CIRs): spiral structures formed by the interaction of gas streams
flowing from the stellar surface with different velocities (Mullan, 1986).
The formation of CIRs has been studied hydrodynamically in 2-D by Cranmer
\& Owocki, (1996) and Lobel \& Blomme (2008), and in 3-D by Dessart (2004).
The 2-D calculations also modelled the effect of the CIRs on the emergent
line profiles, showing that they can reproduce the typical migrating
behaviour of the DACs.  They find that the optical depth enhancements are
essentially caused by a {\em velocity} kink discontinuity slowly migrating
outward in a fixed stellar frame.

More pertinent to the present study is that the spectroscopic end-products of the
evolution of either CIRs or large scale clumps, are localised narrow
absorption components (NACs), which are commonly seen in single (snapshot)
UV spectrograms (e.g. Lamers et al. 1982; Howarth {\&} Prinja
1989).
The definition of a NACs adopted here is that of
narrow absorption
components that occur near the terminal velocity.  While their
strengths
may change with time, they tend to remain fixed is velocity.  Our study
addresses
the information that can be determined from the NACs, while 
acknowledging that DAC formation
and propagation is far more complex
(e.g. Brown et al. 2004)
and beyond the scope of the current work.
The surveys cited above suggest that NACs may be present, at some time, in
most OB stars with well developed but unsaturated wind lines.  For example,
a preliminary examination of the 106 B0-B5 supergiant sample analysed by
Prinja et al. (2005) indicates that NACs are evident in $\sim$70{\%} of
the stars which exhibit well-developed, unsaturated P Cygni profiles in
either Si{\sc iv} or C{\sc iv}.

\begin{table}[t]
\caption{B supergiant sample with decoupled Si{\sc IV} profiles
and clear NACs.
The spectral type to $T_{\rm eff}$ and luminosity
calibration has been adopted from Searle et al. (2008);
parameters for the SMC star AV 264 are from
Trundle {\&} Lennon (2005).
}
\label{log1}
\begin{tabular}{llll}
\hline
\hline
\multicolumn{1}{l}{Star}
&\multicolumn{1}{l}{Sp. type}
&\multicolumn{1}{l}{$T_{\rm eff}$ (kK)}
&\multicolumn{1}{l}{{$v_\infty$}/Si{\sc iv} $\Delta_{\rm sep}$}
\\
\hline
\\

HD 13866    & B2 Ib-II:p  & 18.1 & 0.44 \\
HD 47240    & B1 Ib  & 21.7 & 0.50 \\
HD 111990    & B2 Ib  & 18.1 & 0.37 \\
HD 51309    & B3 Ib  & 17.5 & 0.38 \\
HD 36371    & B4 Ib  & 17.4 & 0.22 \\
AV 264      & B1 Ia & 22.5 & 0.38 \\
\\
HD 91316     & B1 Iab & 21.9 & 0.57 \\
HD 191877    & B1 Ib & 21.7 & 0.55 \\
HD 93840     & BN1 Ib & 21.7 & 0.55 \\
HD 100276    & B0.5 Ib  & 25.4 & 0.65 \\
HD 152234    & B0.5 Ia & 24.7 & 0.66 \\
HD 224151    & B0.5 II & 27.5 & 0.66 \\

\hline
\end{tabular}
\end{table}

\section{The special case of NACs in decoupled Si{\sc iv} $\lambda\lambda$1394,1403 lines}

The general problem with measuring mass-loss rates from UV wind lines is
that the wind line variability we see is due to structures in the wind, so
we can never be certain whether we are seeing a part of the wind with a
covering factor of 1 (the doublet ratios verify this).  On the other hand,
NACs, almost by definition, are parts of the line profiles with the largest
optical depth.  They are formed in a region of the wind where the velocity
law flattens out as it approaches the terminal velocity.  As a result, all
structure at radii greater than the region where the velocity is $\sim$
0.9$\times$ terminal velocity ($v_\infty$) contribute to a very small part
of velocity space.  Consequently, the wind becomes "smooth" on the Sobolev
length at high velocity, so column densities are uniformly distributed over
the stellar disk.  This requires the structures contained within a Sobolev
length to cover the star, but the chances are high for that since the
Sobolev length is so large.  Consequently, if the covering factor is unity
anywhere, it is near terminal velocity, hence the NACs.

In their study demonstrating wide-spread clumping in B supergiant winds,
Prinja {\&} Massa (2010) exploited Si{\sc iv} 1400 resonance line doublets
of B0 to B5 supergiants where the red and blue components of the doublet
are decoupled and formed essentially independently for targets with
relatively low wind terminal velocities (i.e. $v_\infty$ less than the
doublet separation).
The key advantage of exploiting radiatively uncoupled
 doublets
is that these transitions arise from the same level
of the same ion. 
Consequently, they sample exactly the same material.
Therefore regardless of the total optical depth, 
and provided the basic assumptions of the Sobolev
approximation are satisfied,
the ratio of the optical depths of the 
doublets
{\it must} be the ratio of their oscillator strengths
($\sim$ 2 for the Si IV 1400 doublet) unless the
covering factor of the stellar disk is less than unity
(see also Ganguly et al. 1999).
As explained by Prinja {\&} Massa (2010) if the ratio of the
Si{\sc iv} blue-to-red
doublet component optical depth is 
less than 2 at a
given velocity, the covering factor of the stellar surface by the
intervening wind material
at that velocity must be less than one. 
This
applies regardless of how the column density
is accumulated, i.e., whether
by encountering a velocity plateau or a density enhancement.

We acknowledge however that there are two scenarios where in this
context the Sobolev approximation may not be appropriate:
Firstly, the density (or velocity) structure of the wind over a Sobolev length
may be more complex than the Sobolev approximation assumes.  In that case,
a more sophisticated
radiative transfer code would indeed be needed to perform the
calculation, including a detailed prescription of the density/velocity
structure (which in unknown). 
Alternatively, the surfaces of constant velocity in the wind could be porous over a
Sobolev length.
This is indeed the interpretation adopted in the present paper.
One argument that favours this
conjecture is that the absorption components of wind line profiles show
extremely strong temporal variability while their emission components
remain relatively stable.  This can be naturally explained in terms of
covering factors. Further, more sophisticated theoretical investigations
seem to verify this notion Sundqvist et al. (2010).
Of course both of these alternatives may be present
and the wind may be complicated
by a variety of velocity plateaus and density
variations.
Nevertheless for a wind with
complex density and velocity structure in the radial
direction, one should expect similar structures in the
lateral directions.  And this lateral structure can give rise
to variable covering factors less than one.

Thus
we can interpret a change in the apparent doublet ratio as a function of
velocity
as a change in the clumping. 
Furthermore,  when the
doublet ratio is less than the ratio of the
oscillator strengths, the inferred optical depth
is always less than the
actual optical depth. 
This means that derived parameters, such as
the mass
loss rate times the ionization fraction,
\Mdot\,$q$, is always under estimated. 
In their line-synthesis analyses Prinja {\&} Massa (2010) reveal that the mean
ratio of $\tau_{\rm{rad}}(w)$ of the blue to red Si{\sc iv} components are
rarely close to the canonical value of $\sim$ 2 (expected from atomic
constants), and instead trend toward values closer to $\sim$1 and much less
than 2.  These results may be interpreted in terms of a photosphere that is
partially obscured by optically thick structures in the outflowing gas.

We re-visit here the B supergiant sample of Prinja {\&} Massa (2010) to
measure the radial optical depths of the NACs separately for
decoupled blue and red Si{\sc iv} doublet components. We test the above
notion that though the signs of clumping vary throughout the wind, the
(blue-to-red) ratios of the optical depths near the NACs are closer to 2:1
(i.e. ratio of the oscillator strengths) and thus not so strongly affected
by porosity effects.

\begin{figure*} 
\includegraphics[scale=0.5]{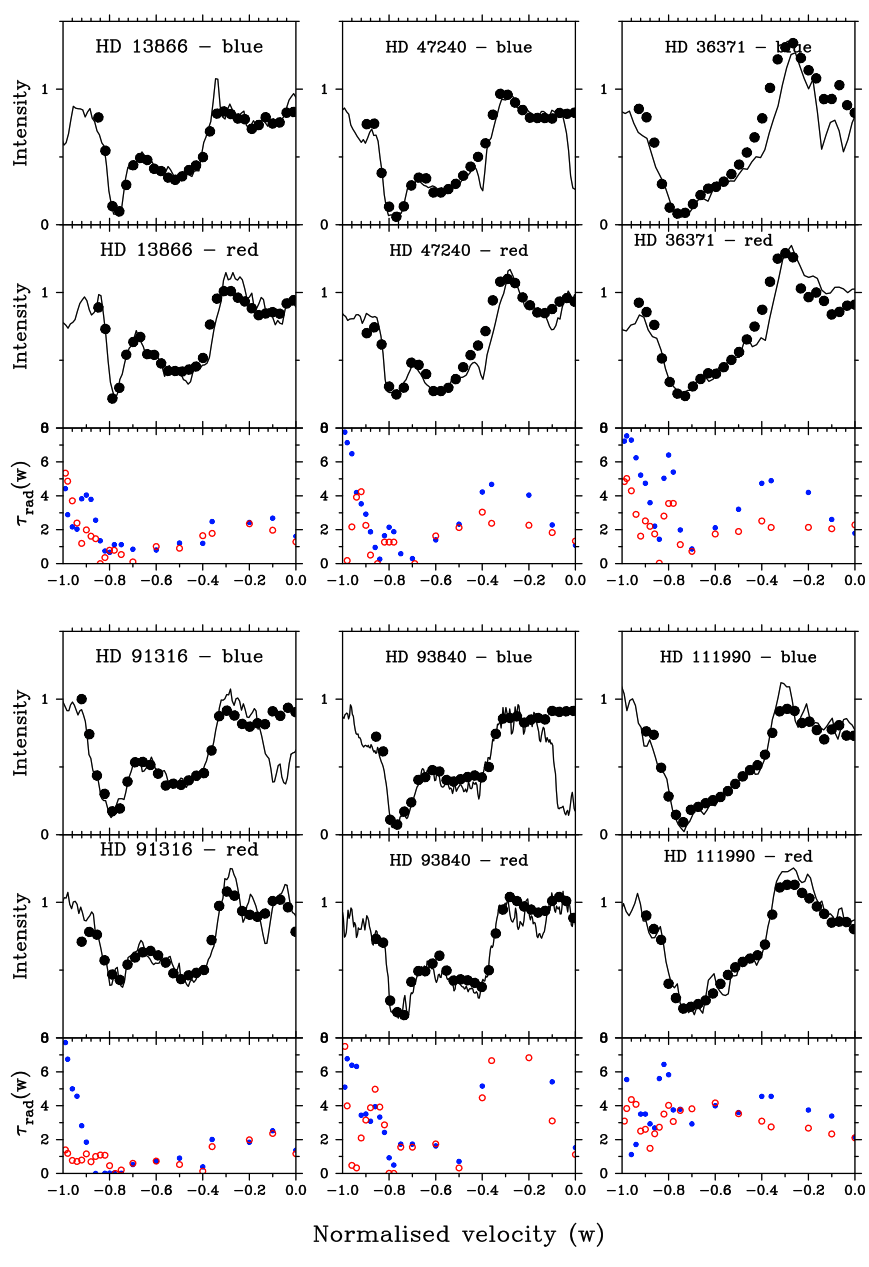} 
\caption{Examples of 
SEI line synthesis fits
(dots) to Si{\sc iv} 
$\lambda$1393.76 (upper panels)
and Si{\sc iv} $\lambda$1402.77 (middle panels) 
for our sample of B supergiants (Table 1).
The lower-most panels show the corresponding radial optical
depths for the blue (blue filled circles) and red (red open circles)
doublet components. (Normalised velocity, w, = v/v$_\infty$)}
\end{figure*}

\section{NAC radial optical depths}
Our primary selection criteria were, firstly, to restrict the
B supergiant sample to stars with $v_\infty$ $\simlt$ 970 km s$^{-1}$
so that the two components of the Si{\sc iv} $\lambda\lambda$1393.76, 1402.77
doublet can be treated as radiatively decoupled, since the radiation
from one component cannot interact with the radiation field of the other
(see e.g. Prinja {\&} Massa 2010). This sample of high-resolution
$IUE$ spectra was then further restricted to stars that exhibit clearly
defined, well-developed NACs close to $v_\infty$. We rejected cases of
very deep NACs which would not allow the radial optical depth ($\tau_{\rm{rad}}(w)$)
to be reliably constrained. We also rejected stars
with very narrow (full width at half maximum less than 0.05 $v_\infty$) NACs
since the velocity width of these features
is less than the resolution of the optical depth bins used
in the line-synthesis; see below.
Unfortunately only 6 B supergiants fit these criteria (see Table 1),
so in order to increase the sample size we relaxed the criteria slightly
to allow additional cases where the $v_\infty$ is
$\simlt$ 0.65 of the Si{\sc iv} doublet separation. Thus an additional
6 stars with good NAC examples were added (Table 1). Though strictly
the doublets in these additional stars can interact, this interaction
is with the diffuse, scattered component due to the blue component
from the far side of the star, which is still weak at the red NAC
velocity and will not introduce a substantial error to the optical depths.

As in our previous studies (e.g. Massa et al. 2003) we derived the radial
optical depths from the observed Si{\sc iv} line profiles by using
the 
Sobolev with Exact Integration
(SEI) method (Lamers et al. 1997), modified to allow the $\tau_{\rm{rad}}(w)$
values to be
determined using a histogram of 21 variable bins in velocity space.
We note that Bouret et al. (2012) have recently demonstrated
that the SEI method is extremely accurate for wind profiles.
The SEI approach offers two key advantages for the present
study:
(i) It is free of assumptions of how the wind material or its
ionization state are distributed, thus making it an ideal tool for empirical
investigations;
(ii) It is extremely fast, so that it can be used in non-linear least
squares routines, where profiles must be computed hundreds of times.

Our approach is to analyse as large a number of
stars as possible and then examine the resulting parameters to search for
systematics.  Such a goal precludes the use of more sophisticated models.
Rotationally broadened $IUE$ Si{\sc iv} spectra of B dwarfs
were used as the photospheric input to the SEI models. (A TLUSTY 
plane-parallel model atmosphere spectra (e.g. Hubeny {\&}
Lanz, 1995)
was adopted for the SMC star AV264.) 
We assume a standard ``$\beta$-law'' for the
expansion of the wind, and set (for $w$ = $v/v_\infty$)
the initial velocity $W_0$ = 0.01
and the extent of the optical depth for resonance lines
$W_1$ = 1.
A non-linear least squares procedure is used to match the wind profiles,
with the velocity law index ($\beta$) and turbulent
velocity parameter ($v_{\rm turb}$) also allowed to vary.
Generally in order to match the NACs well there was a tendency
for $v_\infty$ to be placed very close to the shortward edge of the profile
with low values of $v_{\rm turb}$. The adopted SEI model fitting parameters
are listed in Table 2.

Examples of the independent SEI line profile matches to red and blue
Si{\sc iv} doublet components are shown in Fig. 1, where we also plot
the corresponding values for $\tau_{\rm{rad}}(w)$.  The models generally
match the observed absorption very well for the NAC region and at low
velocities (typically down to $\sim$ 0.3 $v_\infty$).  The velocity ranges
adopted for the individual NACs are listed in Table 3, and in most cases
the NACs are measured in at least 3 velocity bins; nevertheless the
standard deviation in $\tau_{\rm{rad}}(w)$ can be quite large.  The NAC
blue-to-red velocity averaged optical depth ratio, $\langle \tau_{\rm
{rad}}^{\rm blue}/\tau_{\rm {rad}}^{\rm red}\rangle_w$, is also given in
Table 3 where it is compared to the corresponding ratio measured at lower
velocity ($\sim$ 0.4 $-$ 0.6 $v_\infty$).  In all 12 B supergiant stars we
note that $\langle \tau_{\rm {rad}}^{\rm blue}/\tau_{\rm {rad}}^{\rm red}
\rangle_w$ is higher (and closer to $\sim$ 2.0) in the NAC than at lower
velocities (see Fig. 2).  The difference in $\langle \tau_{\rm {rad}}^{\rm
blue}/\tau_{\rm {rad}}^{\rm red} \rangle_w$ between the NACs and low
velocity wind demonstrates two key points; (i) the signs of clumping vary
as a function of velocity throughout the wind and that the stellar disk
covering factor is relatively low for most of the outflow, and (ii) the ratio
of optical depths in the NACs is closer to $\sim$ 2:1 i.e. the ratio of
$f$-value. The covering factor is thus $\sim$ 1 at the (high) NAC velocity.
{\it This suggests that the optical depths of the NACs provide
the most reliable measurement of mass-loss rate in the UV lines.}
Equally, we can picture that {\it doublet} SEI fits to stars in our sample
would result in excellent fits to {\it both} NACs with a single
set of unsaturated $\tau_{\rm {rad}}$ values. However in this case, when
the doublet model matches the blue component at low velocity, the model
will be too shallow (in this region) for the red component.  This is the
tell-tale signature of porosity affecting the lower wind velocity more
severely than the near-$v_\infty$ NAC.

\begin{table*}[t]
\caption{SEI model parameters and optical depth ratios}
\label{log1}
\begin{tabular}{lcccccc}
\hline
\hline
\multicolumn{1}{l}{Star}
&\multicolumn{1}{l}{$v_\infty$ (km s$^{-1}$)}
&\multicolumn{1}{l}{$v_{\rm turb}/v_\infty$}
&\multicolumn{1}{l}{$\beta$}
&\multicolumn{1}{l}{NAC vel. range}
&\multicolumn{1}{l}{$<$$\tau_{\rm {rad}}(w)$(blue/red)$>$ (NAC)}
&\multicolumn{1}{l}{$<$$\tau_{\rm {rad}}(w)$(blue/red)$>$ (low vel.)}

\\
\hline
\\
HD 13866   & 860 & 0.06 & 3.0 & 0.88 $-$ 0.95 & 1.92 $\pm$ 0.40 & 1.07 $\pm$ 0.61 \\
HD 47240  & 970  & 0.12 & 3.0 & 0.85 $-$ 0.98 & 1.75 $\pm$ 0.39 & 1.17 $\pm$ 0.36 \\
HD 111990 & 715 & 0.12 & 3.0 & 0.80 $-$ 0.85 & 1.74 $\pm$ 0.20 & 1.12 $\pm$ 0.19 \\
HD 51309  & 730 & 0.09 & 0.5 & 0.88 $-$ 0.98 & 1.70 $\pm$ 0.86 & 1.34 $\pm$ 0.60  \\
HD 36371  & 465 & 0.16 & 3.0 & 0.75 $-$ 0.85 & 1.83 $\pm$ 0.53 & 1.41 $\pm$ 0.72 \\
AV 264    & 730 & 0.09 & 3.0 & 0.75 $-$ 0.85 & 2.45 $\pm$ 0.44 & 1.69 $\pm$ 0.56 \\

\\

HD 91316  &  1105 & 0.15 & 2.0 & 0.85 $-$ 0.95 & 3.54 $\pm$ 0.50 & 1.30 $\pm$ 0.27 \\
HD 191877 &  1140 & 0.10 & 0.5 & 0.85 $-$ 0.95 & 1.88 $\pm$ 0.65 & 1.45 $\pm$ 0.29 \\
HD 93840  &  1075 & 0.07 & 0.5 & 0.85 $-$ 0.97 & 1.79 $\pm$ 0.63 & 1.02 $\pm$ 0.10 \\ 
HD 100276 &  1260 & 0.16 & 0.5 & 0.83 $-$ 0.93 & 1.92 $\pm$ 0.74 & 1.53 $\pm$ 0.88 \\
HD 152234 &  1275 & 0.05 & 0.8 & 0.95 $-$ 0.98 & 2.97 $\pm$ 1.08 & 1.55 $\pm$ 0.33 \\
HD 224151 &  1280 & 0.05 & 1.5 & 0.85 $-$ 0.95 & 1.76 $\pm$ 0.51 & 1.65 $\pm$ 0.44 \\

\hline
\end{tabular}

\end{table*}

\section{Discussion}

When employing UV wind lines as mass loss diagnostics, only unsaturated 
wind lines are useful.   However, Prinja \& Massa (2010) have shown that 
the optical depths of apparently unsaturated wind lines may be strongly 
affected by porosity, causing mass loss rates derived from UV wind lines 
to under estimate the actual mass loss.  Sundqvist et al. (2010), and 
\v{S}urlan et al. (2012) have presented theoretical models for specific 
wind structure geometry which demonstrate how this effect can occur.  In 
this paper, we have attempted to quantify exactly how large the effects of 
porosity on optical depths can be.

By considering the decoupled Si{\sc iv} $\lambda\lambda$1400 blue and red
doublet components of B supergiants with relatively low terminal velocity
(i.e. $v_\infty$/Si{\sc iv} $\Delta_{\rm sep}$ $\simlt$ 0.65), we have
demonstrated that the blue-to-red optical depth ratios
in NACs are closer
to $\sim$ 2 (as expected from atomic constants) than at lower wind
velocities.  The results indicate that while at most of the wind velocities
optically thick clumps cover only a fraction of the stellar surface,
in the near-$v_\infty$ NAC region the covering factor is $\sim$ unity.
This implies that the NACs provide a better empirical UV measure of
mass-loss rates that are not seriously affected by porosity.

\begin{figure}
\includegraphics[scale=0.2]{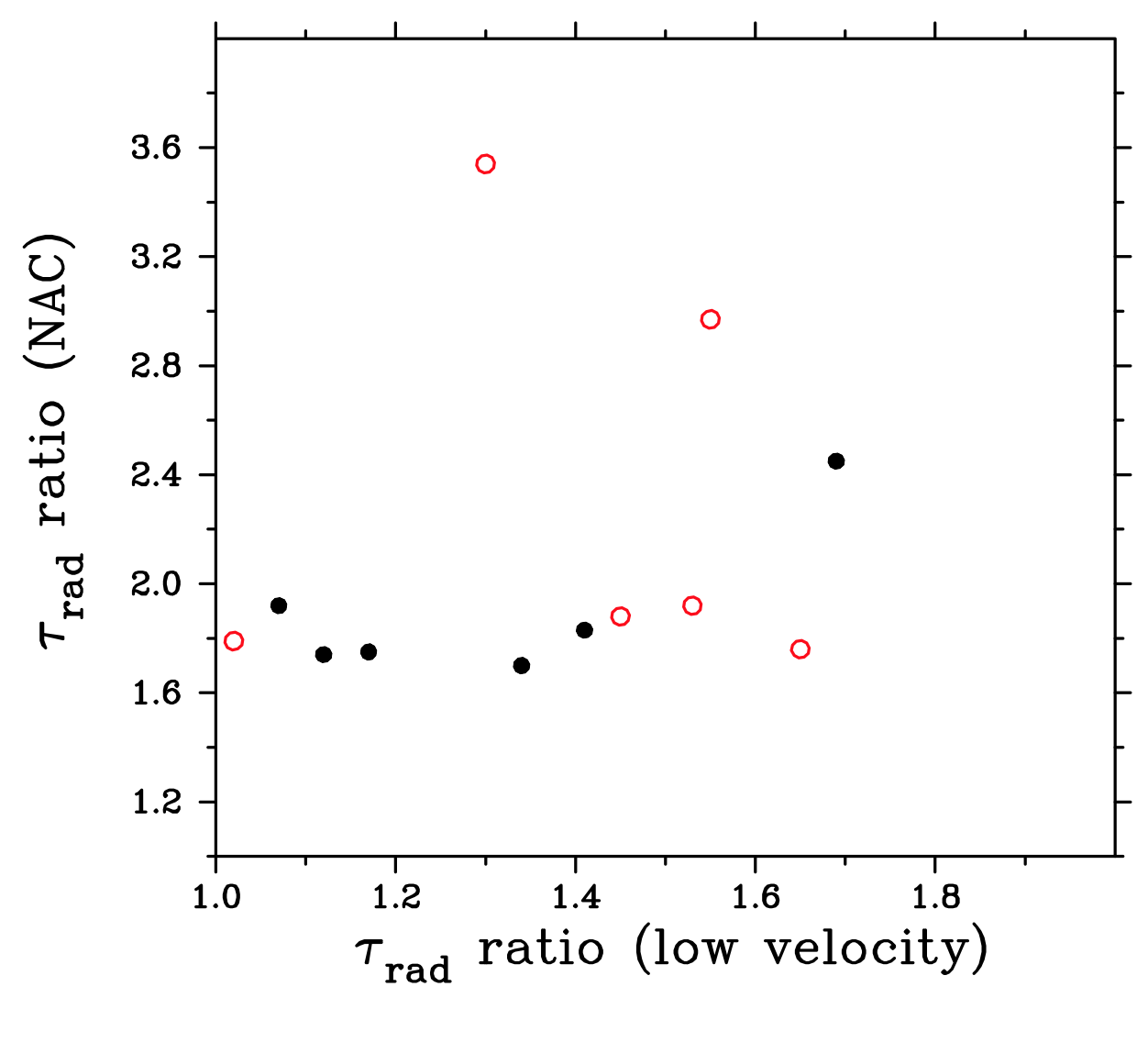}
\caption{The blue-to-red radial optical depth ratio
in the NACs
plotted against the corresponding ratio measured
at low wind velocity ($\sim$ 0.4$-$0.6 $v_\infty$).
Stars with $v_\infty$/Si{\sc iv} $\Delta_{\rm sep}$
$\simlt$ 0.5 and $\simlt$ 0.65 are plotted
in black closed and red open circles, respectively.
}
\end{figure}

The SEI model fitting parameters in Table 2 were used to calculate the
product of mass-loss rate and ion fraction, \Mdot\,$q($Si$^{3+})$.  Mean
values determined over the NAC velocity region are listed in Table 3 for
our sample stars. These estimates may be compared to \Mdot $q($Si$^{3+})$
calculated at lower velocities ($\sim$ 0.4 to 0.6 $v_\infty$), and we also
provide in Table 3 the mass-loss rates obtained from the model 
prescriptions of Vink et al. (2000, 2001), $\dot{M}({\rm Vink})$.  Changes 
of $T_{\rm eff}$ or $L/L_\odot$ resulting from a $\pm$ one spectral or 
luminosity bin error, would change these mass-loss rate predictions by less 
than a factor of 2.

It is clear that \Mdot $q($Si$^{3+})$ is higher when measured in the NACs;
in the 'poster-child' cases of HD~13866 and HD~47240 (which have excellent
NAC profiles) \Mdot $q($Si$^{3+})$ is larger by factors of $\sim$ 9 and 5,
respectively.  Further, recall that some stars in our initial sample were 
discarded because the NACs were saturated, so these values should be 
considered lower limits.  

There is a second consequence of porosity that also affects derived mass 
loss rates.  Whether NACs are due to an accumulation of clumps (Sundqvist 
et al. 2010, \v{S}urlan et al.\ 2012) or CIR arms (see Fig.\ 12 in 
Fullerton et al.\ 1997) in velocity space, both models involve 
inhomogeneous winds.  This makes interpretation of the optical depths of 
NACs in terms of $\dot{M}$ more complex.  For a homogeneous wind, one can 
argue that for some combination of stellar parameters, the ion fraction of 
a specific element, $q$, must approach 1.  For an inhomogeneous wind, this 
is no longer the case.  Different ions of the same element may attain 
their largest ion fractions in different constituents of the wind for 
different stellar parameters.  For example, it is possible that when 
$q($Si$^{2+}) \sim 1$ in the denser wind structures, $q($Si$^{3+}) 
\sim 1$ in the surrounding medium, and that when $q($Si$^{3+}) \sim 1$ in 
the structures, $q($Si$^{4+}) \sim 1$ (which lacks an observable wind 
line) in the surrounding medium.  As a result, the maximum value of $q$ 
for a given ion may only be $\sim 1/2$ or less.  However, constraints can 
be applied to the possible ion stages present.  For example, if a wind 
lacks strong C~{\sc iv} and N~{\sc v} lines, then most of the silicon must 
be in stages Si$^{3+}$ and below, all of which have observable resonance 
lines.

Returning to HD~13866 and HD~47240, their effective temperatures 
approximately correspond to the regime where $q($Si$^{3+}$) peaks (e.g. 
Prinja et al. 2005) and Si$^{3+}$ may be expected to be dominant.  However, 
the spectra of both stars have Si~{\sc iii}$\lambda1206$ and C~{\sc 
iv}$\lambda\lambda 1550$ wind lines that are unsaturated at intermediate 
velocities and have saturated NACs.  
Combining the relative oscillator strengths and
adunances of these lines suggests that Si~{\sc iii}$\lambda1206$ and
C~{\sc iv}$\lambda\lambda 1550$
would be 2.8 and 3.9 times stronger (respectively)
than Si~{\sc iv}$\lambda\lambda 
1400$ for the same column density of wind material.  Thus, it is possible 
that no more than $\sim 1/3$ of the silicon is in Si$^{3+}$ and that even 
though \Mdot $q($Si$^{3+})$ in the NACs of these stars are only $\sim 0.01$ 
$\dot{M}({\rm Vink})$ (Table 3), the actual mass loss rates could easily 
be 3 times larger.  However, that is still a factor of 30 below theoretical 
expectations.  

The UV mass-loss diagnostics discussed here can have wider applications
in the (majority) cases where the resonance line doublets cannot be 
treated as radiatively decoupled. We can adopt the B supergiant results 
derived here and model just the blueward NAC in any well-defined OB star 
unsaturated P Cygni profile.  This is feasible with our methods because we 
are fitting the outer region of the wind where, over a fixed velocity 
increment, the blue NAC region is relatively homogeneous and decoupled from 
the rest of the wind.  An important example is the NACs seen in the 
C{\sc iv} $\lambda\lambda$1550 and N{\sc v} $\lambda\lambda$1240 lines of
weak-wind main-sequence late O-type stars (e.g. Kaper et al. 1996).  The
mass-loss determinations in these weak-wind cases remain a challenge for
the line-driven wind theory (e.g. Bouret et al. 2005).

\begin{table}[t]
\caption{
Comparison of \Mdot $q($Si$^{3+})$  values for NAC and low-velocity
regions of the Si{\sc iv} profile.
\Mdot predictions obtained from Vink et al. (2000, 2001) mass-loss recipes
are also listed. (Z $\sim$ 0.2 Z$_\odot$ was adopted for
the SMC star AV 264).}
\label{log1}
\begin{tabular}{llll}
\hline
\hline
\multicolumn{1}{l}{Star}
&\multicolumn{1}{l}{\Mdot\,$q(Si^{3+})$}
&\multicolumn{1}{l}{\Mdot\,$q(Si^{3+})$}
&\multicolumn{1}{l}{$\dot{M}({\rm Vink})$}
\\

&(blue NAC) & (blue low-vel) & 
\\

 & (10$^{-9}$ M$_\odot$ yr$^{-1}$) & (10$^{-9}$ M$_\odot$ yr$^{-1}$) & ( 10$^{-6}$ M$_\odot$ yr$^{-1}$) 
\\
\hline
\\

HD 13866    & 10.61 & 1.21 & 1.13 \\
HD 47240    & 15.82 & 3.40 & 1.82 \\
HD 111990    & 6.35 & 1.95 & 1.35 \\
HD 51309    & 1.88 & 2.25  & 1.10 \\
HD 36371    & 2.60 & 0.89 & 1.42 \\
AV 264      & 3.10 & 2.41 & 10.50 \\
\\
HD 91316     & 5.69 & 1.32 & 1.81 \\
HD 191877    & 4.06 & 1.29 & 1.55 \\
HD 93840     & 3.26 & 1.09 & 1.65 \\
HD 100276    & 2.38 & 1.94 & 3.54 \\
HD 152234    & 3.04 & 2.43 & 3.40 \\
HD 224151    & 2.07 & 1.17 & 0.17 \\

\hline
\end{tabular}
\end{table}
%

\begin{figure}
\includegraphics[scale=0.30]{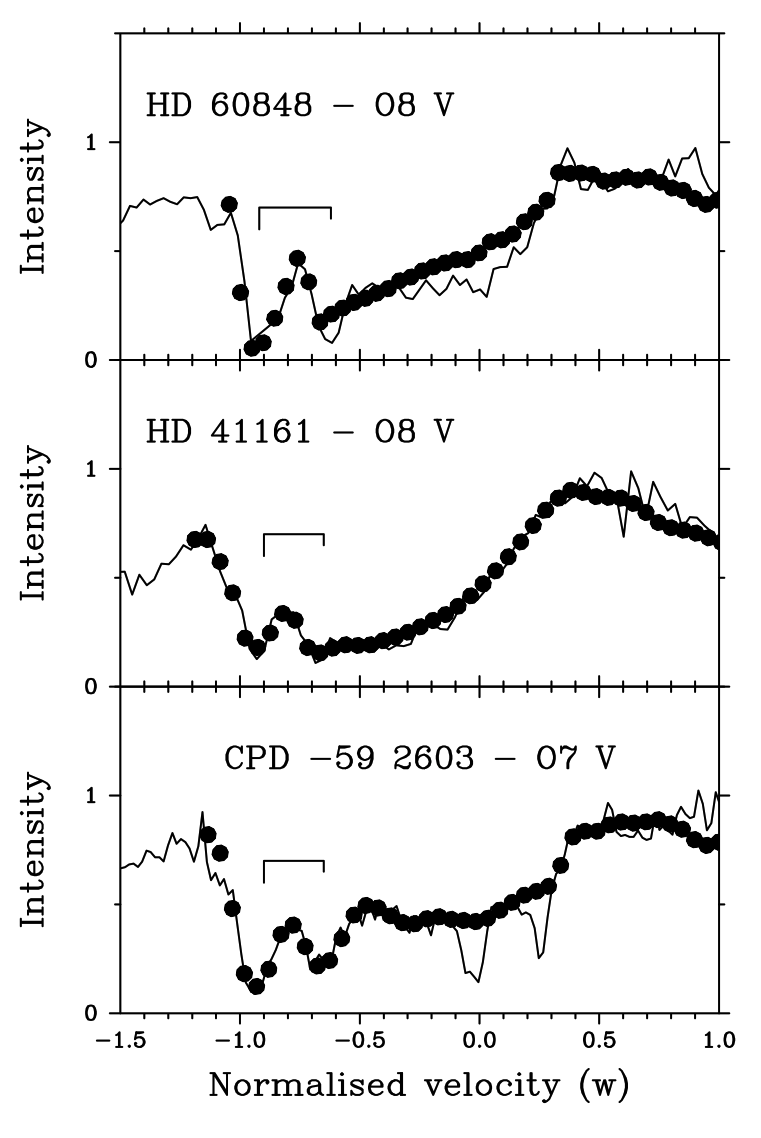}
\caption{Examples of SEI fits to the weak C{\sc iv} $\lambda\lambda$1550
P Cygni doublet profile of O dwarf stars that have clearly developed
blueward NACs
(marked above each doublet component).
(TLUSTY models were used as the photospheric
input in these cases.)
}
\end{figure}

To demonstrate this application to O dwarf stars we fitted the C{\sc iv}
$\lambda\lambda$1550 $IUE$ line profiles of HD~60848 (O8 V pe var),
HD~41161 (O8 V) and CPD -59\,2603 (O7 V((f))). In each of these cases a
well defined, relatively broad, NAC close to $v_\infty$ in weak P Cygni
profiles is evident in $IUE$ high-resolution spectra. The SEI model fits are
shown in Fig. 3. Adopting from the present study that the blue NAC forms 
at a velocity where the stellar photosphere covering factor is $\sim$ 1, 
the \Mdot\,$q($C$^{3+})$ values obtained over the NAC region ($\sim$ 0.9 
to 0.97 $v_\infty$) are $\sim$ 5.5 $\times$ 10$^{-10}$ M$_\odot$ yr$^{-1}$ 
(HD~60848), $\sim$ 1.2 $\times$ 10$^{-9}$ M$_\odot$ yr$^{-1}$ (HD~41161), 
$\sim$ 6.1 $\times$ 10$^{-10}$ M$_\odot$ yr$^{-1}$ (CPD~-59\,2603).  These 
values are a factor of $\sim$ 3 larger than \Mdot\,$q($C$^{3+})$ 
calculated over most of the profile (i.e. $\sim$ 0.2 to 0.9 $v_\infty$).  
An understanding of clumping as a function of wind velocity is clearly 
requisite for the determination of reliable mass-loss rate estimates.  The 
UV analysis relies additionally on a description of the ion fraction 
which, as discussed earlier and by Sundqvist et al. (2010) or \v{S}urlan 
et al.\ (2012), can also be affected by large scale clumping or the 
presence of extremely hot wind plasma (Huenemoerder et al., 2010).  

Our NAC results may also help explain the ``P{\sc v} discordance'' in O 
stars.  Fullerton et al. (2006) presented a compelling, independent 
indication of strong clumping in the wind based on the analysis of the 
P~Cygni P{\sc v} resonance line doublet in {\it FUSE} far-UV spectra.  
Because phosphorus has a low cosmic abundance, this doublet never 
saturates in normal OB stars, providing useful estimates of mass-loss rate 
when P$^{4+}$ is the dominant ion. These mass-loss rates are however 
considerably discordant with those inferred from other diagnostics such as 
H$\alpha$ and free-free emission.  The most reasonable way to resolve this 
difference is to invoke extreme clumping in the wind.  The Galactic O 
supergiants examined by Fullerton et al.  (2006) and also the LMC stars 
studied by Massa et al. (2003) generally do {\it not} exhibit NACs in 
their unsaturated P{\sc v} doublet lines.  This absence of NACs may be 
explained by the ratio of ion fractions derived by Massa et al. (2003) 
which suggest that the massive winds in LMC O stars recombine at large 
radius.  Stars with massive enough winds to have P~{\sc v} at low to 
intermediate velocity, recombine to P~{\sc iv} at high velocity, thus 
we do not see the NACs.  Inspection of the O supergiant (P{\sc v}) sample 
suggests however that several of the Si~{\sc iv} profiles in these stars do 
have NACs.  For stars in this sample with well-developed NACs that clearly 
contrast against the 'underlying' P~Cygni absorption, it would be 
worthwhile to derive \Mdot\,$q($Si$^{3+}$) values for the blue Si{\sc iv} 
NAC since we have demonstrated here that the contrast between \Mdot from 
NACs and from intermediate velocity can be of order 3 to 10. A 
corresponding upward adjustment of the P~{\sc v} \Mdot estimates would 
then place these values closer to the H$\alpha$ estimates and the 
theoretical predictions.

However, when addressing all of these issues in hot star winds, one must 
keep in mind that geometry matters.  Although some progress has been made 
modelling the effects that of randomly distributed  clumps of various 
shapes have on wind lines, there is compelling evidence (e.g.
Kaper et al. 1996; Fullerton et al. 1997; Prinja et al. 2002) that the 
structure in OB winds is not random, but rather the result of CIRs.  
Exactly how CIR structures affect wind lines, X-rays, free-free emission 
and other wind diagnostics is, as yet, largely unexplored.

\begin{acknowledgements}
Support for program HST-GO-12218 was provided by NASA
through a grant from the Space Telescope Science Institute, which is
operated by the Association of Universities for Research in Astronomy,
Inc., under NASA contract NAS 5-26555.
We thank the referee for commenting on the manuscript.
\end{acknowledgements}

{}

\end{document}